# Superconductivity in $CeO_{1-x}F_xFeAs$ with upper critical field of 94 T

**J. Prakash[a], S. J. Singh[b], S. Patnaik[b*] and A. K. Ganguli[a*]**

[a]Department of Chemistry, Indian Institute of Technology, New Delhi 110016 India

[b]School of Physical Sciences, Jawaharlal Nehru University, New Delhi 110067 India

---

[*]**Author for correspondence**

[a]Email: - ashok@chemistry.iitd.ernet.in

Tel No. 91-11-26591511

Fax 91-11-26854715

[a]Email: - spatnaik@mail.jnu.ac.in

## Abstract

We have successfully synthesized Ce based oxypnictide with fluorine doping ($CeO_{1-x}F_xFeAs$) by a two step solid state reaction method. Detailed XRD and EDX confirm the crystal structure and chemical compositions. We observe that an extremely high $H_{c2}$ (0) of 94 T can be achieved in the x = 0.1 composition. This increase in $H_{c2}$ (0) is accompanied by a decrease in transition temperature (38.4 K in x = 0.1 composition) from 42.5 K for the x = 0.2 phase. The in-plane Ginzburg-Landau coherence length is estimated to be ~ 27 Å at x = 0.2 suggesting a moderate anisotropy in this class of superconductors. The Seebeck coefficient confirms the majority carrier to be electrons and strong dominance of electron-electron correlations in this multiband superconductor.

PACS 74.10. +v Occurrence, potential candidates

PACS 74.70.-b Superconducting Materials

PACS 74.62.-c Transition temperature variation

Keywords: Oxypnictides, Superconductivity

**Introduction:**

The recent discovery of superconductivity in LnOFeAs (Ln= Rare earth) type layered compounds (also called oxypnictides) gives a fresh incentive to understand high temperature superconductivity. The parent compound LaOFeAs itself is not superconducting but shows a spin density correlation ~ 150 K while doping with fluorine suppresses the magnetic instability and leads to superconductivity at 26 K [1]. The superconducting state is believed to originate from the quasi two-dimensional Fe-As layers in which Fe forms a square lattice. Both LnOFeAs based superconductors and the cuprates show a strong sensitivity of transition temperature ($T_c$) on the relative atomic position within the transition metal layers. In this case the Fe-As layers provide the conducting pathway and La-O layers act as charge reservoirs [2]. Following the initial work, several other phases were synthesized with $T_c$ as high as 55 K by replacing La with other rare earth elements such as Ce, Sm, Nd, Pr and Gd [3-7]. It has been reported that electron-doped cerium iron oxypnictide superconductors has many features similar to LaOFeAs superconductor [3]. The transition temperature in $CeO_{0.84}F_{0.16}FeAs$ iron-based compound is reported to be 41 K for 16% fluoride doping [3]. Another group has reported a $T_c$ around 30 K for 10 % fluoride doping in CeOFeAs system which increases to around 40 K for 20 % doping with an upper critical field of 48.8 Tesla [8]. Here, we report the magnetic and transport properties of $CeO_{1-x}F_xFeAs$ for x = 0.1 and 0.2 compositions. A $T_c$ of 42.5 K for $CeO_{0.8}F_{0.2}FeAs$ and a substantial enhancement of the upper critical field ($H_{c2}$) to ~94 T for $CeO_{0.9}F_{0.1}FeAs$ are observed. Such high critical field would be very important for several applications.

## Experimental

Polycrystalline samples with nominal compositions of $CeO_{1-x}F_xFeAs$ with x= 0.10 and x = 0.20 were synthesized by a two step solid state reaction method [9-10] using high purity Ce, $CeO_2$, $CeF_3$ and FeAs as starting materials. FeAs was obtained by reacting Fe chips and As powder at 900 ºC for 24 hours. The raw materials (all with purities better than 99.9 %) were taken according to stoichiometric ratio and then sealed in evacuated silica ampoules ($10^{-4}$ torr) and heated at 1000 ºC for 48 hours. The powder was compacted (5 tonnes) and the disks were wrapped in Ta foil, sealed in evacuated silica ampoules and heated at 1180 ºC for 48 hours. All chemicals were handled in a nitrogen filled glove box. The resulting samples were characterized by powder x-ray diffraction method (XRD) with Cu-K α radiation. The x-ray diffraction pattern could be indexed satisfactorily on the basis of the tetragonal ZrCuSiAs-type structure with space group P4/nmm.

Resistivity measurements were carried out using a cryogenic 8 T cryogen-free magnet in conjunction with a variable temperature insert (VTI). This system can access temperature down to 1.6 K and field up to 8 Tesla. In this system the samples were cooled in helium vapor and the temperature of the sample was measured with an accuracy of 0.05 K using a calibrated Cernox sensor wired to a Lakeshore 340 temperature controller. Standard four probe technique was used for transport measurements. Contacts were made using 44 gauge copper wires with air drying conducting silver paste. The external magnetic field (0-5 T) was applied perpendicular to the probe current direction and the data were recorded during the warming cycle with heating rate of 1 K/min. The inductive part of the magnetic susceptibility was measured via a tunnel diode based rf penetration depth technique [11]. The sample was kept inside an inductor that formed a part of an LC circuit of an ultrastable oscillator (~2.3MHz). A change in the magnetic state of the sample results in a change in the inductance of the coil and is reflected as a shift in oscillator frequency which is measured by an Agilent 53131A counter. To obtain the superconducting fraction, DC magnetization studies were carried out using a Quantum Design MPMS SQUID magnetometer. Energy dispersive analysis by X- rays (EDX) was carried out on sintered pellets using a Zeiss electron microscope in conjunction with a BRUKER EDX system. The thermoelectric power data were obtained in the bridge geometry across a 2 mm by 3 mm rectangular disk.

**Results and discussion:**

Figure 1 shows the XRD pattern for the sample with nominal composition of $CeO_{0.9}F_{0.1}FeAs$ and $CeO_{0.8}F_{0.2}FeAs$. All the observed reflections could be satisfactorily indexed based on the tetragonal CeOFeAs (space group *P*4/*nmm*) phase for the x = 0.1 composition (pure phase). Small amount of CeAs and $Fe_2As$ were observed as secondary phases for the x = 0.2 composition. The refined lattice parameters were found to be **a** = 3.991(1) Å and **c** = 8.613(3) Å for x= 0.10 and **a** = 3.988(3) Å and **c** = 8.607(8) Å for the x= 0.2 phase. The lattice parameters are smaller than the parent compound CeOFeAs (a = 3.996 Å and c = 8.648 Å) reported earlier [3]. Compared to the undoped phase (CeOFeAs), the reduction of the lattice volume upon F-doping indicates a successful chemical substitution. In figure 1(c) we present the energy dispersive X-ray microanalysis (EDX) spectrum of one typical grain, which shows that the main elements of the grains are Ce, Fe, As, O and F. The spectrum also confirmed 1:1:1 atomic percentage ratio between Ce, Fe and As. It is thus safe to conclude that the superconductivity observed here comes from the main phase $CeO_{1-x}F_xFeAs$. There is a small peak of Si possibly due to the formation of SiO vapor from the quartz tube employed for the high temperature synthesis.

The inset of figure 2(b) shows the variation of resistivity of $CeO_{0.8}F_{0.2}FeAs$ with temperature. The shape of the curve indicates a good metallic behavior in the normal state. As shown in the main panel of figure 2a and 2b, for $CeO_{0.9}F_{0.1}FeAs$ (sample **a**) and $CeO_{0.8}F_{0.2}FeAs$ (sample **b**), the onset temperatures of the superconducting transition are found to be 38.4 and 42.5 K respectively. The transition width is estimated to be ~ 2 K for sample **a** and ~7 K for sample **b**. The onset temperature for $CeO_{0.8}F_{0.2}FeAs$ is higher than that reported earlier [3, 8], but we observe that the transition is relatively broad. Note that we have used a similar criterion (as elucidated schematically in the figure 2) for determination of transition temperature as has been used elsewhere [9]. We also point out that using the criteria as mentioned in ref. 3 would yield a still higher $T_c$. The residual resistivity ratio (RRR = $\rho_{300}$ / $\rho_{45}$) for $CeO_{0.8}F_{0.2}FeAs$ and $CeO_{0.9}F_{0.1}FeAs$ is about 6.67 and 4.34 respectively. This indicates good intergrain connectivity. The normalized $\chi'$ plotted in the insets of figure 2a and 2b attests the onset of bulk diamagnetic behavior at 38.2 K for sample **a** and 40.4 K for sample **b** respectively. As expected, in these polycrystalline superconductors, magnetic $T_c$ is lower than the resistive $T_c$ and this difference increases with the width of the resistive transition. Nevertheless, we emphasize that there are clear indications that the optimized $T_c$ for $CeO_{0.8}F_{0.2}FeAs$ could be higher than that reported in ref 3. We have also estimated the superconducting volume fraction (~75%) from the DC magnetic susceptibility data (inset of figure 2a) of sample **a**.

The temperature dependence of resistivity under different magnetic fields for (a) $CeO_{0.9}F_{0.1}FeAs$ and (b) $CeO_{0.8}F_{0.2}FeAs$ are shown in figure 3. It is clear that the onset temperature shifts gradually with magnetic field while the zero resistance temperature varies significantly. Near $T_c$, the data show considerable flux-flow broadening as shown in figure 3 where the magnetic field is varied from 0 to 4 T. The transition width indicates a broad region of flux-flow resistivity of possible thermally activated origin near the offset. By using the criteria of 90% and 10% of normal state resistivity ($\rho_n$), we can calculate the upper critical field ($H_{c2}$) and the field H*(T), the so called irreversibility field in High $T_c$ cuprate parlance. By using the Werthamer-Helfand-Hohenberg (WHH) formula [12], the zero field upper critical field $H_{c2}$ (0) can be calculated as well; $H_{C2} = -0.693 T_C \left( {dH_{C2}}/{dT} \right)_{T=T_C}$. The slope of $dH_{c2}/dT$ estimated from the $H_{c2}$ versus T plots are -3.52 and -1.45 T/K for samples **a** and **b** respectively. Using the transition temperature of $T_c$ = 38.4 K and 42.5 K, we find $H_{c2}$ (0) = 94 T and 43 T for $CeO_{0.9}F_{0.1}FeAs$ and $CeO_{0.8}F_{0.2}FeAs$ respectively. This value is approximately twice as compared to the value reported for $CeO_{0.9}F_{0.1}FeAs$ earlier, $H_{c2}(0)$ = 48.8 T [8]. All our field measurement was carried out with the magnetic field perpendicular to the surface of the sample. Evidently, we are able to significantly increase upper critical field with dilute alloying which is a clear signature of impurity scattering in multiband scenario [13]. Using the value of $H_{c2}$ (0) we can also calculate the mean-field Ginzburg-Landau coherence length by the formula $\xi_{ab} = \left( {\phi_0}/{2\pi H_{C2}} \right)^{1/2}$. Taking $\Phi_0 = 2.07 \times 10^{-7}$ G cm$^2$ and the $H_{c2}$ (0) values, we obtain a coherence length of ~ 19 Å and ~ 27 Å for sample **a** and **b** respectively. These results show the moderate anisotropic nature of this class of superconductors as compared to extremely anisotropic High - $T_c$ cuprates. Further, to get more insight into the conduction mechanism, we have obtained the Seebeck coefficient from the temperature dependent thermoelectric power measurement (inset of figure 3b). At room temperature, the coefficient is negative (S = - 21 μV/K) and shows a minimum at 64 K (S = - 104 μV/K). The magnitude corresponding to minimum is slightly higher compared to that reported for La(O/F)FeAs [9]. By comparing with the Mott expression $S = \pi^2 k_B T (2eT_F)^{-1}$, and taking the coefficient at 40 K (that extrapolates to zero corresponding to 0 K), we estimate that the observed thermopower is almost an order of magnitude higher than the free electron prediction. This reconfirms strong electron – electron correlations in the series of oxyarsenide superconductors. The value of Seebeck coefficient remains negative for all temperatures above the transition temperature ($T_c$), which suggests electron to be the predominant charge carriers as have been suggested from Hall measurement [8].

**Conclusion**

We have obtained cerium based oxypnictide superconductors, $CeO_{1-x}F_xFeAs$ for x = 0.10 and 0.20 having the $T_c$ of 38.4 and 42.5 K respectively. This $T_c$ is the highest reported so far in Ce -based oxypnictides. A significant increase in upper critical field along with lower $T_c$ and sharper transition are achieved with 10% F doped sample. Temperature dependent resistivity measurements in the presence of magnetic field show robust flux pinning in these superconductors. These features can potentially make the Ce based oxypnictides much more attractive for potential applications as compared to $MgB_2$ and low $T_c$ Nb-based industrial superconductors.

**Acknowledgement**

AKG and SBP thank DST, Govt. of India for financial support. JP and SJS thank CSIR and UGC, Govt. of India, respectively for fellowships. SBP thanks AIF, JNU for the EDX measurements.

## Figure captions

**Figure: 1.** Powder X-ray diffraction patterns (XRD) patterns of (a) $CeO_{0.9}F_{0.1}FeAs$ and (b) $CeO_{0.8}F_{0.2}FeAs$ annealed at 1180 °C. The impurity phases are $Fe_2As$ (*) and CeAs (O) and (c) A representative EDX spectra of $CeO_{0.9}F_{0.1}FeAs$.

**Figure: 2.** Temperature dependence of normal state resistivity (ρ) as a function of temperature for (a) $CeO_{0.9}F_{0.1}FeAs$ and (b) $CeO_{0.8}F_{0.2}FeAs$ samples under zero magnetic field. Upper inset of figure (a) shows the inductive part of susceptibility as a function of temperature and lower inset shows the temperature dependence of magnetic susceptibility in the zero-field cooled (ZFC) and field-cooled (FC) cycles at 100 Oe for sample **a**. Upper inset of figure (b) shows resistivity plot (till 290 K) and lower inset shows temperature dependent normalized magnetic susceptibility for $CeO_{0.8}F_{0.2}FeAs$.

**Figure: 3.** Temperature dependence of the electrical resistivity of (a) $CeO_{0.9}F_{0.1}FeAs$ and (b) $CeO_{0.8}F_{0.2}FeAs$ in magnetic field (applied perpendicular to the probe current). Inset of figure (a) shows temperature dependence of the upper critical field (■) and irreversibility field (●) as a function of temperature and inset of figure (b) shows temperature dependence of Seebeck coefficient for $CeO_{0.9}F_{0.1}FeAs$ in zero magnetic field.

**Figure1.**

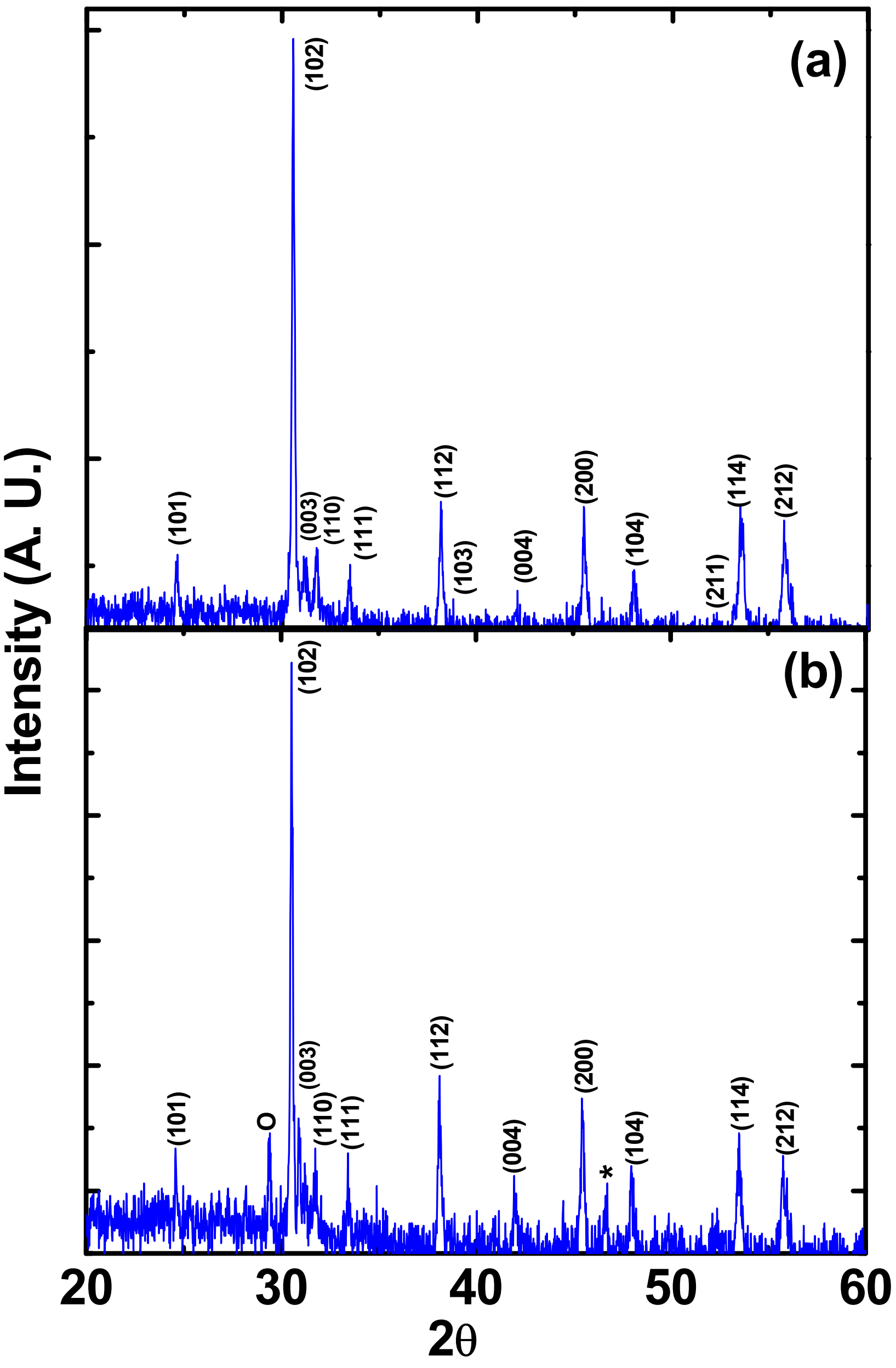

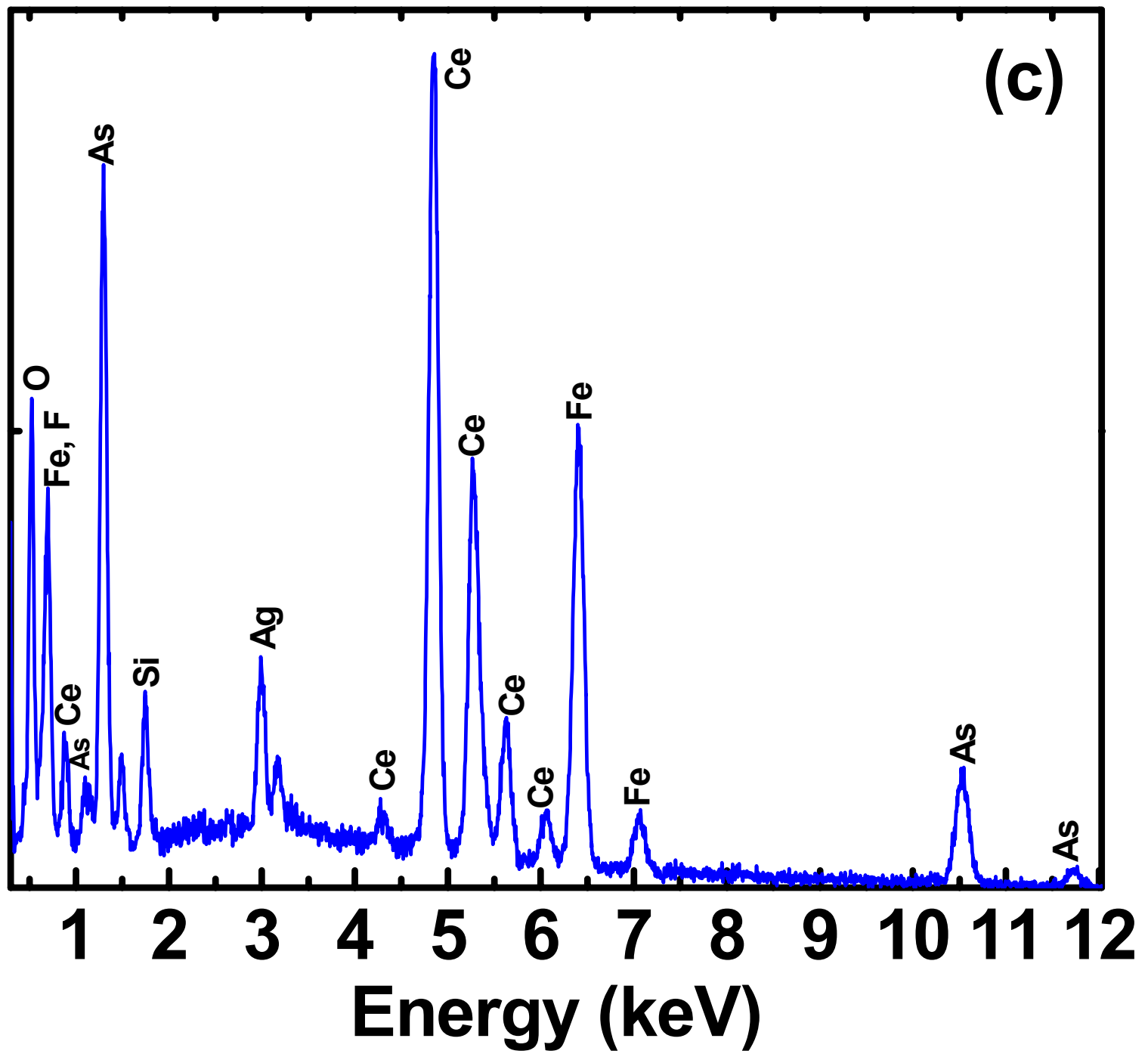

Figure 2.

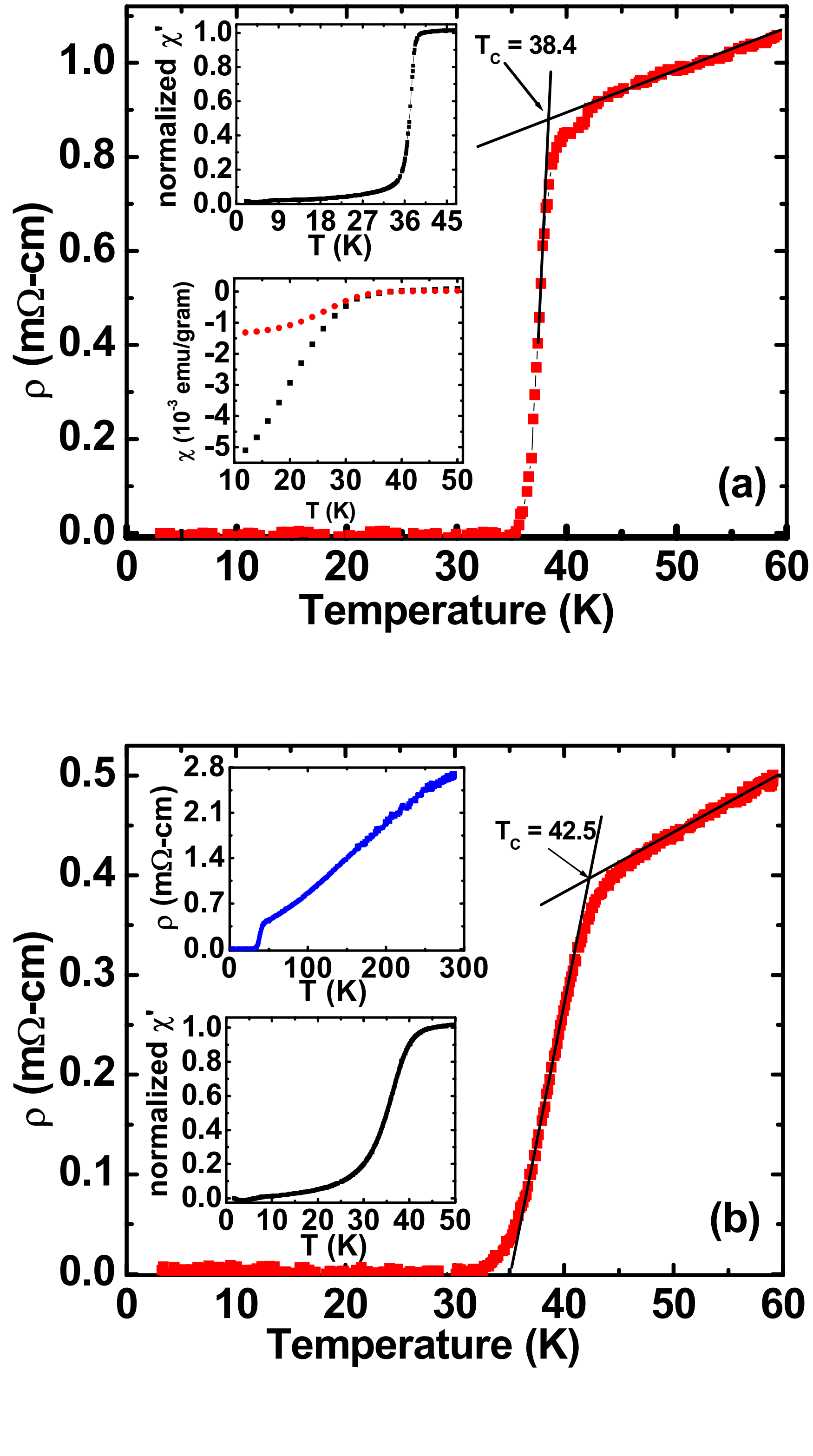

**Figure 3.**

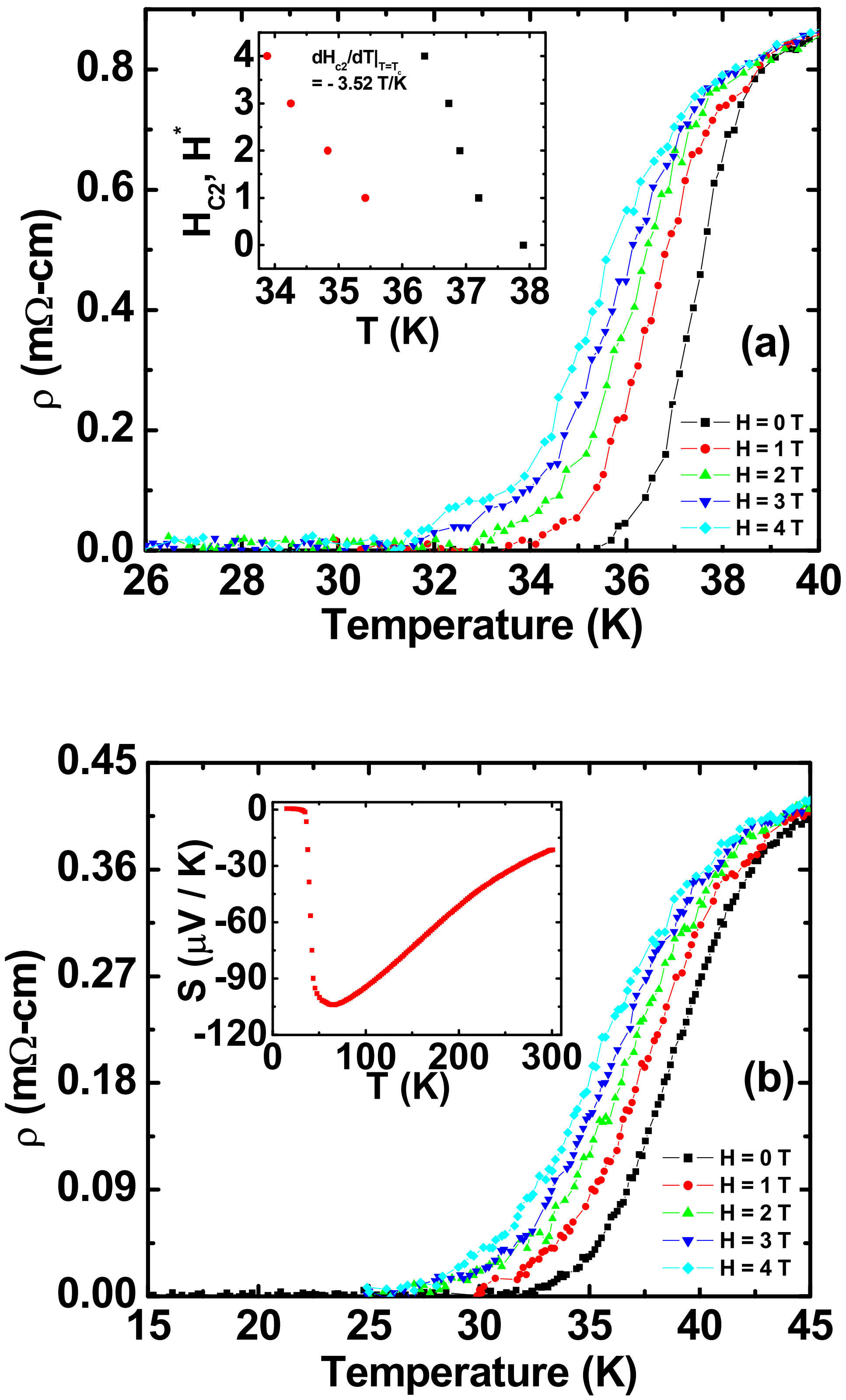

(a)
ρ (mΩ-cm)
Temperature (K)
0.0
0.2
0.4
0.6
0.8
26
28
30
32
34
36
38
40
H = 0 T
H = 1 T
H = 2 T
H = 3 T
H = 4 T
$H_{C2}$, $H^*$
T (K)
$dH_{c2}/dT|_{T=T_c}$ = - 3.52 T/K
(b)
ρ (mΩ-cm)
Temperature (K)
0.00
0.09
0.18
0.27
0.36
0.45
15
20
25
30
35
40
45
H = 0 T
H = 1 T
H = 2 T
H = 3 T
H = 4 T
S (μV / K)
T (K)
0
-30
-60
-90
-120
100
200
300